\documentclass[12pt]{article}

\renewcommand{\d}{{\rm d}}
\newcommand{\M}{{\cal M}}
\newcommand{\R}{{\cal R}}
\newcommand{\D}{{\cal D}}
\newcommand{\B}{{\cal B}}
\newcommand{\N}{{\cal N}}
\newcommand{\F}{{\cal F}}
\newcommand{\arsh}{\mathop{\rm arsh}\nolimits}
\newcommand{\sh}{\mathop{\rm sh}\nolimits}
\begin{document}

\title{Simplicial Palatini action
}

\author{V.M. Khatsymovsky \\
 {\em Budker Institute of Nuclear Physics} \\ {\em of Siberian Branch Russian Academy of Sciences} \\ {\em
 Novosibirsk,
 630090,
 Russia}
\\ {\em E-mail address: khatsym@gmail.com}}
\date{}
\maketitle

\begin{abstract}
We consider the piecewise flat spacetime and a simplicial analog of the Palatini form of the general relativity (GR) action where the discrete Christoffel symbols are given on the tetrahedra as variables that are independent of the metric. Excluding these variables classically gives exactly the Regge action.

This paper continues our previous work. Now we include the parity violation term and the analogue of the Barbero-Immirzi parameter introduced in the orthogonal connection form of GR. We consider the path integral and the functional integration over connection. The result of the latter (for certain limiting cases of some parameters) is compared with the earlier found result of the functional integration over connection for the analogous {\it orthogonal} connection representation of Regge action.

These results, mainly as some measures on the lengths/areas, are discussed for the possibility of the diagram technique where the perturbative diagrams for the Regge action calculated using the measure obtained are finite. This finiteness is due to these measures providing elementary lengths being mostly bounded and separated from zero, just as finiteness of a theory on a lattice with an analogous probability distribution of spacings.
\end{abstract}

keywords: Einstein theory of gravity; piecewise flat spacetime; Regge calculus; affine connection; Palatini action; Christoffel symbol


MSC classes: 83C27; 53C05

\section{Introduction}

The discrete approach is one of the standard tools in analysing such an essentially nonlinear field theory as GR. In particular, it may be of interest for quantizing this theory \cite{Ham} because of the non-renormalizability of the original continuum gravity from the standpoint of the standard criteria. Of the various discrete forms of gravitation, the formulation of the mini-superspace can be of particular interest. It is the GR theory on a certain subclass of the Riemannian spacetime, large enough to be able to approximate in a certain topology any given Riemannian spacetime with arbitrarily high accuracy. This subclass is specified as the piecewise flat spacetime, which can be viewed as a curved spacetime composed of the flat 4-simplices (4-dimensional tetrahedra) or simplicial complex, and the GR theory for this subclass is known as the Regge calculus \cite{Regge}; see also review in \cite{RegWil}. The geometry and GR action for this spacetime is completely described by the set of the edge lengths of this manifold. Another approach related to the Regge calculus, Causal Dynamical Triangulations, has led to important results in quantum gravity \cite{cdt}.

The minisuperspace provides in a sense maximally fine discretization of the theory. Using minisuperspace formulation allows to work with countable number of the degrees of freedom at the same time maximally keeping exact simple relations typical for the continuum theory. One of the types of such relations are some first order representations of the Einstein action, which themselves simplify the theory. The set of independent variables in the Regge calculus, the edge lengths, can be extended by adding some connection type variables, so that the action takes a simple form, but it reduces to the original Regge action after excluding the new variables using the classical equations of motion. By analogy with the continuous theory, in discrete gravity this is also called first-order formulation in the literature and is studied, examples are as follows. The discrete connection and curvature, which are the finite orthogonal rotations introduced in this formulation, were considered in \cite{Fro}. Finite orthogonal rotations were used in the Hamiltonian analysis of the Regge calculus in \cite{Ban3}. Using the discrete analogs of the connection and curvature \cite{Fro}, we have found an exact representation of the Regge action \cite{our1}, a discrete version of the Cartan-Weyl representation of the Einstein action with the orthogonal connection. An approximate version was considered in \cite{CasDadMag}. Another approach was considered in \cite{Bar}, where the interior dihedral angles of simplexes and the areas of triangles are treated as independent variables.

The use of the orthogonal connection involves introducing a set of the local inertial frames. These frames are naturally attributed to the 4-simplices. The same edge has different representations as a 4-vector in the different 4-simplexes containing this edge (correspondingly, an additional constraint is needed restricting these 4-vectors to have the same length). In contrast, the use of the affine connection implies one piecewise affine world frame. This frame is completely determined by specifying the coordinates of the vertices. The complete set of variables consists of the edge lengths, the coordinates of the vertices, and general linear nonsingular 4$\times $4 matrices (elements of GL(4,R) group) on the tetrahedra. We have proposed this formulation in \cite{our2}.

In the present paper we generalize this formulation to the parity violation term, an analog of the Holst term in the Cartan-Weyl formulation (Section \ref{act}). In the case of the finite rotations, it is convenient to choose certain curvature matrices as independent variables instead of a part of the connection matrices. In addition, with this choice it is convenient to use a value of the type of a scale of discrete lapse-shift functions as a parameter, which in principle can be taken small. This is considered in Section \ref{indep}. In Section \ref{path} we consider the path integral and the functional integration over connection. Thus, we obtain its form on the variables of purely metric or lengths. Estimates of it in some simple limiting cases of parameters are made. In the concluding Section, we discuss them together with the earlier found analogous result for the orthogonal connection. In these cases, the pure module of the resulting expression appears in the leading order in the scale of discrete lapse-shift functions. An alternative expansion is a perturbation-theory expansion based on the Taylor series over connection of the connection form of the action around the classical solution, that is, around Regge action. This exists at small lapse-shifts as well, and this and the above expansion over the scale of lapse-shifts have compatible convergence areas. This expansion leads to the pure phase (Regge action) in the leading order and is convenient to represent the latter. The resulting form of the pure metric/length path integral (which sets the rule of quantum averaging of an observable by functional integrating it being multiplied by the weight whose module and phase are just being discussed) is discussed from the viewpoint of the possibility of a finite diagram technique.

\section{The action}\label{act}

We use the terminology in which the Palatini action is a representation of the Einstein action in terms of the affine connection, and the representation in terms of the orthogonal connection is the Cartan-Weyl form. A more general case is the presence of the Holst term \cite{Holst,Fat} which is parametrized by the Barbero-Immirzi parameter \cite{Barb,Imm}. Similarly, a parity violation term can be added to the Palatini action,
\begin{eqnarray}\label{S(G)}                                                
S = \int g^{\mu \nu} R^{\lambda }{}_{\mu \lambda \nu} (\Gamma ) \sqrt{-g} \d^4 x + \frac{1}{2 \gamma} \int \epsilon^{\lambda \mu \nu \rho} R_{\lambda \mu \nu \rho} (\Gamma ) \d^4 x \nonumber \\
R^{\lambda } {}_{\mu \nu \rho} (\Gamma ) = \partial_\nu \Gamma^\lambda_{\mu \rho} - \partial_\rho \Gamma^\lambda_{\mu \nu} + \Gamma^\lambda_{\sigma \nu} \Gamma^\sigma_{\mu \rho} - \Gamma^\lambda_{\sigma \rho} \Gamma^\sigma_{\mu \nu}, ~~~ \epsilon^{0123} = +1.
\end{eqnarray}

\noindent Here we have introduced the notation $1 / \gamma$ for the coefficient before the parity violating term, where $\gamma$ is the direct analogue of the Barbero-Immirzi parameter. This is because a representation analogous to the Cartan-Weyl representation with the general $\omega^{ab}_\lambda$ (not necessarily $\omega^{ab}_\lambda$ = $ - \omega^{ba}_\lambda$) is obtained from the representation (\ref{S(G)}) with the general $\Gamma^\lambda_{\mu \nu}$ with torsion simply by redefinition of $\Gamma^\lambda_{\mu \nu}$ (as some transformation matrix) from the world reference frame to the locally flat reference frames, $\Gamma^\lambda_{\mu \nu} = \omega^{ab}_\nu e^\lambda_a e_{b \mu} + e^\lambda_a \partial_\nu e^a_\mu, ~~~ e^\lambda_a e_{b \lambda} = \eta_{ab} \equiv {\rm diag} (-1, +1, +1, +1)$.

Note that $\epsilon^{\lambda \mu \nu \rho} R_{\lambda \mu \nu \rho} (\Gamma ) \neq 0$ means that the torsion is nonzero. In the simplicial theory, the symbol $\Gamma^\lambda_{\mu \nu}$ is replaced by some matrix $\M^\lambda_{ \sigma^3 \mu }$ on the tetrahedra $\sigma^3$s, and the symmetry condition $\Gamma^\lambda_{\mu \nu} = \Gamma^\lambda_{\nu \mu}$ is transformed into that condition on the matrices $\M^\lambda_{ \sigma^3 \mu }$ on {\it the different} $\sigma^3$s. The last condition is nonlocal and therefore unnatural. Thus, the torsion and the parity odd term are natural in the simplicial case.

Earlier we have proposed a simplicial analogue of this action at $1 / \gamma = 0$ and in Euclidean notation \cite{our2},
\begin{eqnarray}                                                            
\int g^{\mu \nu} R^{\lambda }{}_{\mu \lambda \nu} (\Gamma ) \sqrt{g} \d^4 x & & \nonumber \\ & & \hspace{-20mm} \Rightarrow 2 \sum_{\sigma^2} A(\sigma^2 ) \arcsin \frac{\left ( \R - \R^{-1} \right )^\lambda {}_\tau (\M ) g^{\tau \mu } \Delta x^\nu_{\sigma^1_1} \Delta x^\rho_{\sigma^1_2} \epsilon_{\lambda \mu \nu \rho} \sqrt{g}}{8 A(\sigma^2 )}
\end{eqnarray}

\noindent where $A(\sigma^2 )$ is the area of the triangle $\sigma^2$, $\sigma^1_1$ and $\sigma^1_2$ are two of the sides of this triangle, $\Delta x_{\sigma^1_i}$ is the difference in the coordinates of the ends of the edges $\sigma^1_i$. The connection matrix $\M_{\sigma^3}$ on the tetrahedron $\sigma^3$ is the general linear nonsingular 4$\times $4 matrix. $\R_{\sigma^2} (\M )$ is the curvature matrix, the holonomy of the connection taken along a closed path enclosing the triangle $\sigma^2$ and passing only through the tetrahedra $\sigma^3$ containing this triangle,
\begin{equation}\label{R=PiM}                                               
\R_{\sigma^2} = \prod_{\sigma^3 \supset \sigma^2} \M^{\epsilon (\sigma^2, \sigma^3)}_{\sigma^3},
\end{equation}

\noindent $\epsilon (\sigma^2, \sigma^3) = \pm 1$ is some sign function. It is assumed that the metric $g_{\lambda \mu}$ is taken in the 4-simplex, in which this path begins and ends.

At $1 / \gamma \neq 0$ and in the notation of Minkowski space-time (locally) we need to make a distinction between the norm of the bivector $V^{\lambda \mu}$ of $\sigma^2$ (the genuine area of $\sigma^2$) and the norm of the dual bivector $v_{\lambda \mu}$ of $\sigma^2$ which differ by $\sqrt{-1}$. The connection and curvature matrices still form the general linear group GL(4,R),
\begin{eqnarray}                                                            
& & S^{\rm discr}_{\rm GL(4,R)} = 2 \sum_{\sigma^2} \sqrt{v_{\sigma^2} \circ v_{\sigma^2}} \arcsin \frac{\left ( \R - \R^{-1} \right )^\lambda {}_\tau (\M ) g^{\tau \mu } v_{\sigma^2 \lambda \mu}}{4 \sqrt{v_{\sigma^2} \circ v_{\sigma^2}}} \nonumber \\
& & + \frac{2}{\gamma} \sum_{\sigma^2} \sqrt{V_{\sigma^2} \circ V_{\sigma^2}} \arcsin \frac{\left ( \R - \R^{-1} \right )^\tau {}_\mu (\M ) g_{\tau \lambda } V^{\lambda \mu}_{\sigma^2 }}{4 \sqrt{V_{\sigma^2} \circ V_{\sigma^2}}}
\end{eqnarray}

\noindent where
\begin{eqnarray}                                                            
& & v_{\sigma^2 \lambda \mu} = \frac{1}{2} \sqrt{- g} \epsilon_{\lambda \mu \nu \rho} \Delta x^\nu_{\sigma^1_1} \Delta x^\rho_{\sigma^1_2}, \nonumber \\
& & V^{\lambda \mu}_{\sigma^2 } = \frac{1}{2} \left ( \Delta x^\lambda_{\sigma^1_1} \Delta x^\mu_{\sigma^1_2} - \Delta x^\mu_{\sigma^1_1} \Delta x^\lambda_{\sigma^1_2} \right ),
\end{eqnarray}

\noindent $\epsilon_{0123} = -1, V \circ V \equiv V_{\lambda \mu} V^{\lambda \mu} /2$ etc.

The equations of motion arising when the connection is varied are obtained by applying the operator $\M^\nu_{ \sigma^3 \lambda } \partial /\partial \M^\nu_{ \sigma^3 \mu }$ to the action,
\begin{eqnarray}                                                            
- \frac{1}{2} \hspace{-10mm} \sum_{ \hspace{10mm} \{ \sigma^2 : ~ \sigma^2 \subset \sigma^3 \} } \hspace{-10mm} \epsilon (\sigma^2, \sigma^3) \Gamma_2 (\sigma^2, \sigma^3) \left [ \frac{v_{\sigma^2} \R^{\epsilon (\sigma^2, \sigma^3)}_{\sigma^2 } + \R^{ - \epsilon (\sigma^2, \sigma^3)}_{\sigma^2 } v_{\sigma^2} }{\cos \alpha (\sigma^2)} \right. & & \nonumber \\ & & \hspace{-55mm} \left. + \frac{V_{\sigma^2} \R^{\epsilon (\sigma^2, \sigma^3)}_{\sigma^2 } + \R^{ - \epsilon (\sigma^2, \sigma^3)}_{\sigma^2 } V_{\sigma^2} }{\gamma \cos \alpha^* (\sigma^2)} \right ] \Gamma^{-1}_2 (\sigma^2, \sigma^3) = 0.
\end{eqnarray}

\noindent Here,
\begin{eqnarray}                                                            
\hspace{-0mm} \alpha (\sigma^2 ) = \arcsin \frac{\left ( \R - \R^{-1} \right )^\lambda {}_\tau (\M ) g^{\tau \mu } v_{\sigma^2 \lambda \mu}}{4 \sqrt{v_{\sigma^2} \circ v_{\sigma^2}}}, \nonumber \\ \alpha^* (\sigma^2 ) = \arcsin \frac{\left ( \R - \R^{-1} \right )^\tau {}_\mu (\M ) g_{\tau \lambda } V^{\lambda \mu}_{\sigma^2 }}{4 \sqrt{V_{\sigma^2} \circ V_{\sigma^2}}}.
\end{eqnarray}

\noindent $\Gamma_1$, $\Gamma_2$ denote the products of the connection matrices that appear in the expression for $\R_{\sigma^2}$ in the form $[\Gamma_1 (\sigma^2, \sigma^3) \M_{\sigma^3} \Gamma_2 (\sigma^2, \sigma^3)]^{\epsilon (\sigma^2, \sigma^3)}$. There is a particular ansatz in which $\M_{\sigma^3}$ is a metric-compatible connection, and $\R_{\sigma^2}$ rotates around $\sigma^2$ by the angle $\alpha (\sigma^2 )$ and in the $\sigma^2$ plane by the angle $\alpha^* (\sigma^2 )$ (one can even choose $\alpha^* (\sigma^2 ) = 0$). Then
\begin{eqnarray}\label{vR+Rv}                                               
v_{\sigma^2} \R^{\epsilon (\sigma^2, \sigma^3)}_{\sigma^2 } + \R^{ - \epsilon (\sigma^2, \sigma^3)}_{\sigma^2 } v_{\sigma^2} = 2 v_{\sigma^2} \cos \alpha (\sigma^2 ), \nonumber \\ V_{\sigma^2} \R^{\epsilon (\sigma^2, \sigma^3)}_{\sigma^2 } + \R^{ - \epsilon (\sigma^2, \sigma^3)}_{\sigma^2 } V_{\sigma^2} = 2 V_{\sigma^2} \cos \alpha^* (\sigma^2),
\end{eqnarray}

\noindent and the equations of motion read
\begin{equation}                                                            
\hspace{-10mm} \sum_{ \hspace{10mm} \{ \sigma^2 : ~ \sigma^2 \subset \sigma^3 \} } \hspace{-10mm} \epsilon (\sigma^2, \sigma^3) \Gamma_2 (\sigma^2, \sigma^3) \left ( v_{\sigma^2} + \frac{1}{\gamma } V_{\sigma^2} \right ) \Gamma^{-1}_2 (\sigma^2, \sigma^3) = 0.
\end{equation}

\noindent This expresses the closure of the sum of the combinations of bivectors $\Gamma_2 (v_{\sigma^2} + V_{\sigma^2} / \gamma ) $ $ \Gamma^{-1}_2$ over the surface of the tetrahedron $\sigma^3$ with $\Gamma_2 (\sigma^2, \sigma^3)$ transporting these bivectors to the same 4-simplex, and this is the identity.

\section{Independent connection variables}\label{indep}

Consider a possibility of using independent curvature matrices as a part of the connection variables. This turns out to be conveniently associated with the possibility of using discrete lapse and shift functions as non-dynamic parameters, which in principle can be made small.

The holonomy of the connection along a path enclosing all the triangles sharing an edge reduces to a product of the curvature matrices on these triangles. At the same time, this equals unity, because this path contracts to a point. In fact, the expressions for the curvatures (\ref{R=PiM}) on the full set of the triangles sharing a given edge $\sigma^1_1$ just solve an algebraical identity (the Bianchi identity \cite{Regge}), so we can take all but one of these matrices as independent. This one of the matrices $\R_{\sigma^2_1}, \sigma^2_1 \supset \sigma^1_1$, is a certain (multiplicative) function of the independent curvature matrices and of some connection matrices taken as independent. We can take another edge $\sigma^1_2 \neq \sigma^1_1$ of this triangle $\sigma^2_1$ and consider the curvatures on the full set of the triangles sharing $\sigma^1_2$. As a result, the curvature on some triangle $\sigma^2_2 \supset \sigma^1_2, \sigma^2_2 \neq \sigma^2_1$, can be considered as a function of $\R_{\sigma^2_1}$ and of the curvatures on the triangles $\sigma^2 \supset \sigma^1_2, \sigma^2 \neq \sigma^2_1, \sigma^2 \neq \sigma^2_2$, and of some connection matrices taken as independent. Continuing the process of expressing the curvature matrices in terms of independent ones, we obtain a chain of triangles $\F$ on which the curvature matrices are functions of independent curvature matrices, and also of some connection matrices taken as independent ones: $\sigma^2_1, \sigma^2_2, ... , \sigma^2_n, ... $. Here any two successive triangles $\sigma^2_{n-1}, \sigma^2_n$ share an edge $\sigma^1_n = \sigma^2_{n-1} \cap \sigma^2_n$, which we can call an "internal" edge; the edges of the triangles of $\F$ which are not internal edges can be called "lateral" edges. This process assumes that $\F$ has not self-intersections (over internal edges), and three triangles of $\F$ can not meet at an edge (an edge can not be internal and lateral simultaneously).

The choice of $\F$ can be illustrated by the combinatorially simplified structure where we use cubes instead of simplices. A 4-dimensional spacetime composed of the 4-cubes can be constructed of a sequence of the 3-dimensional spaces composed of the 3-cubes by connecting by edges corresponding vertices between each two successive such 3-cube lattices. By analogy with the simplicial case, these 3-cube lattices can be called leaves of the foliation. The n-cubes completely contained in the leaf can be called "leaf" n-cubes. These lattices can be numbered by the monotonically growing parameter $t$ which can be taken in the role of the time coordinate $x^0$. The edges connecting the neighboring 3-cube lattices can be called "t-like" edges. Also call the n-cubes containing t-like edges "t-like" n-cubes. If we temporarily use the same notation $\sigma^n$ for the n-cube as for the n-simplex, then we have at each vertex some four edges, t-like $\sigma^1_0$, connecting the vertex with the corresponding one in the neighboring 3-cube lattice, corresponding, for definiteness, to a larger value of t, and 3 edges $\sigma^1_i, i = 1, 2, 3$, in the 3-cube lattice itself. We can take the world vector $\Delta x^\lambda_{\sigma^1_0}$ of $\sigma^1_0$ to have zero components at $\lambda = 1, 2, 3$ (and $\Delta x^0_{\sigma^1_i} = 0$ at $i = 1, 2, 3$ since we have assigned an unambiguous $x^0$ to each 3-cube lattice). Consider for a moment 3+1 splitting of the tetrad and metric by Arnowitt-Deser-Misner \cite{ADM},
\begin{equation}                                                           
\d s^2 = \eta_{ab} (e^a_\lambda \d x^\lambda) (e^b_\mu \d x^\mu) = - (N \d t )^2 + [^{(3)}e^k_\alpha (\d x^\alpha + N^\alpha \d t)]^2
\end{equation}

\noindent where $\eta_{ab} = {\rm diag} (-1,1,1,1)$, $^{(3)}e^k_\alpha$ is 3x3 (space) part of the tetrad $e^a_\lambda$, with lapse $N$ and shift $N^\alpha$ functions. Then the tetrad components of the t-like edge are $e^a_\lambda \Delta x^\lambda_{\sigma^1_0} = (N, ^{(3)}e^k_\alpha N^\alpha ) \Delta x^0_{\sigma^1_0}$. We can consider $N$ also as a scale for $N^\alpha$, that is, $N^\alpha = O(N )$ if $N \to 0$. Then the module of the bivectors $v_{\sigma^2}$, $V_{\sigma^2}$ for t-like $\sigma^2$ as well as their products with the bivectors of the other $\sigma^2$ depend on $N$ through the product $N \Delta x^0_{\sigma^1_0}$. This means that developing a perturbation theory over $\Delta x^\lambda_{\sigma^1_0}$ (considering contribution of the areas built on $\sigma^1_0$ and $\sigma^1_i, i = 1, 2, 3$, into the connection form of the action as a perturbation) we get an expansion over $N$ at the same time. To have correspondence with the continuum theory, where the lapse-shift functions are gauge parameters, we can consider them as set by hand in the discrete framework too. Therefore, $N$ can be a small parameter (at least if the number of cells in time direction is finite and fixed, which is usual in the discrete theory). As for the dependent curvature matrices, analogously to the previous paragraph, these can be curvature matrices on the areas (quadrangles) belonging to the chain of quadrangles $\F$. The latter can be taken to be the whole set of the t-like quadrangles, that is, built on $\sigma^1_0$s and $\sigma^1_i$s, $i = 1, 2, 3$. ($\F$ consists of the components - the bands continuing along the t-like edges $\sigma^1_0$.) This fact is successfully combined with the fact that just the contribution of the t-like quadrangles which is rather cumbersome contribution of the dependent curvatures can be considered as a perturbation with some parameter ($N$) which can be made small.

The set of independent curvature matrices consists of the curvature matrices on the leaf quadrangles. There is one-to-one correspondence between the leaf quadrangles and t-like 3-cubes (3 per vertex) since the latter have the former as their bases. Correspondingly, one can successively (step-by-step in time direction) multiplicatively express the connection on the t-like 3-cubes in terms of the curvature on the leaf quadrangles and thus pass from the former to the latter as new variables. The remaining connection variables are those on the leaf 3-cubes (1 per vertex).

The above is modified only slightly if we consider a real simplicial complex. The simplest periodic one was used in \cite{RocWil}. It is obtained by decomposing 4-cubic periodic cell with the help of all the diagonals drawn from some vertex of the 4-cube into 24 4-simplices. This can be considered as consisting of 3-dimensional simplicial complexes (leaves of the foliation) connected with the help of not only t-like, but also {\it diagonal} edges. In fact, the leaf can be generalized to be quite general 3-dimensional simplicial complex; only the method of connecting the leaves is specified: it is assumed that these have the same structure, corresponding vertices are connected by the edges (t-like edges), thus dividing the spacetime between any two neighboring leaves into the 4-prisms with the 3-simplices as bases, and each 4-prism is divided with the help of diagonals into the minimal number (four) 4-simplices, see Fig. \ref{prism}.
\begin{figure}[h]
\unitlength 1pt
\begin{picture}(150,95)(-100,20)
\put(40,110){\line(6,-1){90}}

\put(51,92){a}

\put(40,80){\line(6,1){40}}
\put(5,76){$t + \Delta t$}
\put(83,85){1}
\put(132,91){$t + \Delta t$}
\put(130,95){\line(-6,-1){40}}

\put(102,70){b}

\put(40,80){\line(5,-2){33}}
\put(75,63){2}
\put(115,50){\line(-5,2){33}}

\put(45,60){c}

\put(25,50){\line(1,0){40}}
\put(20,46){$t$}
\put(68,46){3}
\put(119,46){$t$}
\put(115,50){\line(-1,0){40}}

\put(25,50){\line(1,2){15}}
\put(25,50){\line(1,-2){15}}

\put(45,33){d}

\put(40,20){\line(5,2){33}}
\put(76,32){4}
\put(115,50){\line(-5,-2){33}}

\put(107,30){e}

\put(40,20){\line(1,0){40}}
\put(5,16){$t - \Delta t$}
\put(83,16){5}
\put(132,16){$t - \Delta t$}
\put(130,20){\line(-1,0){40}}

\put(40,80){\line(0,1){30}}
\put(115,50){\line(1,3){15}}
\put(115,50){\line(1,-2){15}}
\end{picture}
\caption{A t-like 3-prism and neighboring leaves $t - \Delta t$, $t$, $t + \Delta t$, ..., 2d picture. The t-like tetrahedrons a, b, c, d, e, ..., leaf (1, 3, 5, ... ) and diagonal (2, 4, ... ) triangles. $\R^{\pm 1}_1 = ... \M^{\pm 1}_{\rm a} ... \M^{\pm 1}_{\rm b} ... $ etc, $\D \M_{\rm a} \D \M_{\rm b} ... \D \M_{\rm e} ... = \D \R_1 \D \R_2 ... \D \R_5 ...$}
\label{prism}
\end{figure}
This illustrates the procedure of passing from (the part of) the connection variables to the independent curvature variables. In addition to the t-like simplices (containing t-like edges) and leaf simplices (completely contained in a leaf), there are also {\it diagonal} simplices not belonging to these two groups. The leaf and diagonal simplices in the considered procedure are regarded on equal footing.

Apply this to the path integral measure. The part of the measure depending on the affine connection is the product over the 3-simplices of the group invariant (Haar) measures,
\begin{equation}                                                           
\prod_{\sigma^3} \D \M_{\sigma^3}, ~~~ \D \M = \frac{\d^{16} \M}{(\det \M)^4}.
\end{equation}

\noindent Taking into account the multiplicative nature of the dependence of the connection on the t-like 3-simplices on the curvature on the leaf/diagonal triangles we have due to the invariance of the measure
\begin{equation}                                                           
\prod_{{\rm t-like} \sigma^3} \D \M_{\sigma^3} = \prod_{{\rm leaf/diagonal} \sigma^2} \D \R_{\sigma^2}
\end{equation}

\noindent and for the full measure
\begin{equation}                                                           
\prod_{\sigma^3} \D \M_{\sigma^3} = \prod_{{\rm leaf/diagonal} \sigma^3} \D \M_{\sigma^3} \prod_{{\rm leaf/diagonal} \sigma^2} \D \R_{\sigma^2}.
\end{equation}

To summarize, the curvature matrices on the leaf/diagonal triangles can be considered as independent, then those on the t-like triangles should be treated as dependent ones. The contributions of the latter enter with the coefficients $N$, the scale of the lapse-shift function and, in principle, can be taken as a perturbation (say, when performing the functional integration over connection), which means the expansion in powers of $N$.

\section{Path integral}\label{path}

Consider integration over connection in the path integral. Upon discussing general properties, we dwell on the zero order $N = 0$ when the action reduces to the sum of independent (as functions of connection) terms over the (leaf/diagonal) triangles.

Consider any such term in the action and the related connection type factor in the path integral measure. If one descends from the side of areas much larger than the Planck scale, the largest contribution into path integral comes from 'arcsin' much less than 1, and 'arcsin' can be substituted by its argument. As a result, the corresponding term in the action reads (with $(16 \pi G)^{-1}$ in the action in usual units)
\begin{equation}                                                           
- \frac{1}{32 \pi G} {\rm tr} [\B (\R - \R^{-1})], ~~~ \B^\lambda {}_\mu = v^\lambda {}_\mu + \frac{1}{\gamma} V^\lambda {}_\mu,
\end{equation}

\noindent and the corresponding factor in the path integral
\begin{equation}\label{f(B)}                                               
\int \exp \left \{ \frac{- i}{32 \pi G} {\rm tr} [\B (\R - \R^{-1}) ] \right \} \D \R \equiv f (\B ).
\end{equation}

\noindent We adopt a view that $f (\B )$ is a distribution to be further integrated for averaging any function of the area tensors, here $\B^\lambda {}_\mu$. To this end, we can integrate $f (\B )$ with all the monomials of the components $\B^\lambda {}_\mu$. We can change the order of integration over $\D \R$ and $\d^{16} \B$, which is a component of the definition of the path integral as a conditionally convergent integral. Thus we get some derivative of the $\delta$-function $\delta^{16} (\R - \R^{-1})$ which being integrated over $\D \R$ leads to some finite answer. (In particular, the integral with any arbitrarily high power of area is finite.) Thereby the result of integration of $f (\B )$ with any polynomial of $\B$ is known and $f (\B )$ itself can be restored.

In the case $N \neq 0$ an analogous situation holds for $\int \! $ $ \! \exp \! $ $ \! (iS^{\rm discr}_{\rm GL(4,R)} ) \! $ $\! \prod_{{\rm leaf/diagonal} \sigma^2} \! $ $ \! \D \R_{\sigma^2 }$ (with 'arcsin' replaced by its argument in $S^{\rm discr}_{\rm GL(4,R)}$) instead of (\ref{f(B)}). We consider this expression as a function of independent $\B^\lambda_{\sigma^2 \mu}$s. We can integrate it with some product of the components $\B^\lambda_{\sigma^2 \mu}$ over {\it finite} number of the leaf/diagonal $\sigma^2$s over $\prod_{{\rm leaf/diagonal} \sigma^2} \d^{16} \B_{\sigma^2}$. This gives some finite answer, a {\it moment} of this expression. Thereby the result of integration of this expression with any polynomial of $\B_{\sigma^2 }$ is known and finite.

Finiteness of the moments of this expression is an argument that this expression falls off at large areas faster than any power of areas, probably exponentially, and singularities at finite areas, if any, are integrable. These moments depend analytically on $N$, also in the neighborhood of $N = 0$, which allows us to expect these properties of the dependence on $N$ also for the expression itself. If after the replacement of arcsin x by x we do an inverse transition, then x is replaced by the faster changing function arcsin x, thereby strengthening the suppression properties of the oscillating exponent and again leading to finite moments and the above properties of the original functional integral over connections.

Due to the invariance with respect to $\B \to \N \B \N^{-1}$, $\N \in {\rm GL(4,R)}$, the function $f (\B )$ can depend on the invariants of $\B$ only; meanwhile, $\det \B = \gamma^{-2} A^4$ is singular at $\gamma^{-1} = 0$ or $\gamma = 0$, and these cases require a separate consideration.

Consider the symmetry transformation of the variables
\begin{eqnarray}                                                           
\Delta x^\lambda \to \Delta \tilde{x}^\lambda = \N^\lambda {}_\mu \Delta x^\mu, ~~~ g_{\lambda \mu} \to \tilde{g}_{\lambda \mu} = g_{\nu \rho} (\N^{-1} )^\nu {}_\lambda (\N^{-1} )^\rho {}_\mu, \nonumber \\ \R^\lambda {}_\mu \to \tilde{\R}^\lambda {}_\mu = \N^\lambda {}_\nu \R^\nu {}_\rho (\N^{-1} )^\rho {}_\mu
\end{eqnarray}

\noindent and choose it so that $g_{\lambda \mu} = \eta_{\lambda \mu} \equiv {\rm diag} (-1, +1, +1, +1)$. Then omit tildes. Such $\N$ is fixed up to any additional possible transformation preserving the metric $\eta_{\lambda \mu}$, that is, from SO(3,1). Choose it so that one of the two area tensors, $v_{\lambda \mu}$ and $V_{\lambda \mu}$, contribution from which in the action is considered as dominating one at the moment, would occupy the block $\lambda, \mu = 0, 1$ (if this tensor is timelike; not to be confused with "t-like") or $\lambda, \mu = 2, 3$ (if this tensor is spacelike). Besides that, it is convenient to choose the order of sequence of indices in the vector or matrix notations as $\lambda, \mu, ... = 0, 1, 2, 3$ or $\lambda, \mu, ... = 1, 2, 3, 0$, respectively, so that this block would be the left upper $2 \times 2$ block, with indices $a, b, c, ... = 0, 1$ or 1, 2. Correspondingly, let us consider $\R$ as a $2 \times 2$ matrix whose elements are in turn $2 \times 2$ matrices.
\begin{equation}                                                           
\R = \left (\begin{array}{cc} X & U \\ W & Y \end{array} \right ),
\end{equation}

\noindent then
\begin{equation}                                                           
\R^{-1} = \left (\begin{array}{cc} (X - U Y^{-1} W)^{-1} & -(X - U Y^{-1} W)^{-1} U Y^{-1} \\ -Y^{-1} W (X - U Y^{-1} W)^{-1} & (Y - W X^{-1} U)^{-1} \end{array} \right )
\end{equation}

\noindent and
\begin{equation}                                                           
\D \R = \frac{\d^4 X \d^4 Y \d^4 U \d^4 W}{(\det Y)^4 [\det (X - U Y^{-1} W)]^4}.
\end{equation}

\noindent We can pass from the four $2 \times 2$ matrix variables $U, W, X, Y$ to the set $U, W, X, Z = X - (X - U Y^{-1} W)^{-1}$. In the measure $\D \R$, we can successively pass from $Y$ to $Q = (U Y^{-1} W)^{-1} = W^{-1} Y U^{-1}$ so that $\d^4 Y = (\det W)^2 (\det U)^2 \d^4 Q$; then pass from $Q$ to $Q^{-1}$ so that $\d^4 (Q^{-1}) = (\det Q)^{-4} \d^4 Q$, and so on; in few steps we find the resulting measure in the new variables being factorized,
\begin{equation}                                                           
\D \R = \frac{\d^4 U}{(\det U)^2} \frac{\d^4 W}{(\det W)^2} \d^4 X \d^4 Z.
\end{equation}

If one descends from the side of $\gamma = \infty$ and $v_{\lambda \mu}$ is spacelike ($V_{\lambda \mu}$ and the area itself, by usual definition, is timelike) denote
\begin{eqnarray}                                                           
\hspace{-0mm} v_{\lambda \mu} = \left (\begin{array}{cc} E_{a b} & 0 \\ 0 & 0 \end{array} \right ), ~~ E_{ab} = \left (\begin{array}{cc} 0 & |A | \\ - |A | & 0 \end{array} \right ), ~~ a, b = 1, 2, ~~ \sqrt{v_{\lambda \mu} v^{\lambda \mu} / 2} = |A |, \nonumber \\ A \equiv \sqrt{V_{\lambda \mu} V^{\lambda \mu} / 2} = i|A |.
\end{eqnarray}

\noindent Besides that, $\d^4 Z = \d^4 (Z^a {}_b ) = \d^4 (Z_{ab} )$, and $Z_{ab}$ can be decomposed into some 3-dimensional symmetric $Z_{\rm S}$ and 1-dimensional antisymmetric $Z_{\rm A}$ parts,
\begin{equation}                                                           
\hspace{-0mm} \d^4 Z = \d Z_{11} \d Z_{22} \d \frac{Z_{12} + Z_{21}}{2} \d (Z_{12} - Z_{21}) \equiv \d^3 Z_{\rm S} \d Z_{\rm A}, ~~~ \d Z_{\rm A} = \d (Z_{12} - Z_{21}),
\end{equation}

\noindent so that only $\d Z_{\rm A}$ integration is nontrivial in the factor in the path integral of interest,
\begin{equation}\label{int-dz}                                             
\int \exp \left ( i \frac{|A |}{8 \pi G} \arcsin \frac{Z_{12} - Z_{21}}{4} \right ) \d (Z_{12} - Z_{21}).
\end{equation}

As mentioned above, we adopt the point of view on the path integral with partially performed integration that it is a distribution to be further integrated for averaging any function of the area. The equation (\ref{int-dz}) can be formally continued as an even function of $x = |A |$ to the whole real axis and integrated with the integer powers of $x$ as probe functions. In such an approach, only local properties at zero argument of 'arcsin' are essential. As was quite expected, the answer is obtained simply by changing the variable $Z_{12} - Z_{21} = 4 \sin \phi$ in (\ref{int-dz}) and integrating over $\phi$ over the whole real axis, $(- \infty , + \infty )$ (that is, not only over the domain of the main branch of the arcsine, $[- \pi / 2 , + \pi / 2 ]$).
\begin{eqnarray}\label{A=8piG}                                             
\hspace{-0mm} \int \exp \left ( \frac{i |A |}{8 \pi G} \arcsin \frac{Z_{12} - Z_{21}}{4} \right ) \d (Z_{12} - Z_{21}) = \int \exp \left ( \frac{i |A |}{8 \pi G} \phi \right ) 4 \cos \phi \d \phi \nonumber \\ \hspace{-25mm} = 2 \int \left [ \exp \left ( \frac{i |A |}{8 \pi G} \phi - i\phi \right ) + \exp \left ( \frac{i |A |}{8 \pi G} \phi + i\phi \right ) \right ] \d \phi \nonumber \\ \hspace{-25mm} = 32 \pi^2 G [ \delta ( |A | - 8 \pi G) + \delta ( |A | + 8 \pi G) ] \Rightarrow 32 \pi^2 G \delta ( |A | - 8 \pi G).
\end{eqnarray}

\noindent This looks as appearance of some additional terms in the effective action. However, the situation is not completely described by some unique effective action; rather, there are effective actions with terms $- \phi $ and $+ \phi $ appearing with the same amplitudes.

If the area is spacelike, $v_{\lambda \mu}$ is timelike,
\begin{eqnarray}                                                           
\hspace{-0mm} v_{\lambda \mu} = \left (\begin{array}{cc} E_{a b} & 0 \\ 0 & 0 \end{array} \right ), ~~ E_{ab} = \left (\begin{array}{cc} 0 & |A | \\ - |A | & 0 \end{array} \right ), ~~ a, b = 0, 1, ~~ \sqrt{v_{\lambda \mu} v^{\lambda \mu} / 2} = i |A |, \nonumber \\ A \equiv \sqrt{V_{\lambda \mu} V^{\lambda \mu} / 2} = |A |.
\end{eqnarray}

\noindent The term in the action of interest contains arcsine of an imaginary value, or arsh,
\begin{equation}                                                           
\frac{i |A |}{8 \pi G} \arcsin \frac{Z_{01} - Z_{10}}{4i} = \frac{ |A |}{8 \pi G} \arsh \frac{Z_{01} - Z_{10}}{4}.
\end{equation}

\noindent Now, after changing the variable $Z_{01} - Z_{10} = 4 \sh \psi$, we see that the path integral is poorly defined; interpretation by means of an effective action would require imaginary terms in it. On the other hand, $\arsh x$ is slowly varying (as $\ln |x |$ at $|x | \to \infty$) function, and the situation is sensitive to any dependence on $Z_{ab}$ through $\R$ of those terms that are neglected here. These terms are the contribution of the t-like triangles mentioned above, which can depend on $\R$ indirectly through the dependent curvatures. The t-like triangles are largely determined by the discrete lapse-shift vectors set by hand, and, as a rule, are timelike. Then their contribution includes arcsine of real arguments. This restricts the possible values of $Z_ {ab}$ (when other variables are fixed) so that these arguments are in the required region $[- 1, +1]$. In addition, if all the arcsine branches should be taken into account, as mentioned in the paragraph after the equation (\ref{int-dz}), this arcsine effectively dominates this logarithm whatever small the coefficient at arcsine (some area) might be.

Analogous situation holds at $\gamma \to 0$ for the timelike area,
\begin{equation}                                                           
\hspace{-2mm} V_{\lambda \mu} = \left (\begin{array}{cc} E_{a b} & 0 \\ 0 & 0 \end{array} \right ), ~~ E_{ab} = \left (\begin{array}{cc} 0 & |A | \\ - |A | & 0 \end{array} \right ), ~~ a, b = 0, 1, ~~ A \equiv \sqrt{V_{\lambda \mu} V^{\lambda \mu} / 2} = i |A |.
\end{equation}

\noindent The term in the action of interest contains arsh, and the contributions that are neglected here can be decisive for integrating over $\D \R$.

As for the spacelike area at small $\gamma$,
\begin{equation}                                                           
\hspace{-1mm} V_{\lambda \mu} = \left (\begin{array}{cc} E_{a b} & 0 \\ 0 & 0 \end{array} \right ), ~~ E_{ab} = \left (\begin{array}{cc} 0 & |A | \\ - |A | & 0 \end{array} \right ), ~~ a, b = 1, 2, ~~ A \equiv \sqrt{V_{\lambda \mu} V^{\lambda \mu} / 2} = |A |.
\end{equation}

\noindent Similarly to the equation (\ref{A=8piG}), we effectively obtain a constraint that fixes this area,
\begin{equation}\label{A=8piGgamma}                                        
\hspace{-0mm} \int \exp \left ( \frac{i |A |}{8 \pi G \gamma} \arcsin \frac{Z_{12} - Z_{21}}{4} \right ) \d (Z_{12} - Z_{21}) \Rightarrow 32 \pi^2 G \delta ( |A | - 8 \pi G \gamma).
\end{equation}

Above we limited ourselves to one term in the action and abstracted from other terms - the contribution of t-like triangles and the contribution of the tensor of the same triangle, but dual-conjugate (another power of $\gamma$ from $\gamma^0$, $\gamma^{-1}$). When these contributions are taken into account, the delta functions obtained above broaden.

Note that if we start from the approximate representation in which arcsin x is replaced by x, then in the equations (\ref{A=8piG}), (\ref{A=8piGgamma}) we simply obtain the delta function $\delta (|A |)$, which equates $A$ to zero.

\section{Discussion and conclusion}

We have considered a simplicial version of the Palatini action for gravity including the parity violating term. We have analyzed the functional integration over the connection part of the path integral measure. Arguments are given that the result as a function of area tensors falls off at infinity faster than any inverse polynomial, its singularities at finite areas (if any) are integrable, the dependence on a scale of the discrete lapse-shift functions $N$ analytical. This can be illustrated by a more detailed analysis in simple limiting cases. Useful is a possibility to expand over $N$ around zero value at which the functional integral over connection is factorized into factors referred to separate triangles. We analyse such a factor as a function of the triangle area. We are able to calculate it in closed form in two limiting (singular) cases, at $\gamma \to \infty$ in the region of the timelike area and at $\gamma \to 0$ in the region of the spacelike area. The evaluated expression takes a simple limiting form $\delta ( | A | - 8 \pi G )$ and $\delta ( | A | - 8 \pi G \gamma )$ for the module of area $|A |$, timelike and spacelike, respectively (at finite nonzero $\gamma$ the $\delta$ function should broaden and become smooth function). Among other things, this emphasizes the role of the parity violating term, despite its zero contribution to the classical equations of motion.

Earlier we have found analogous properties of the path integral for the orthogonal connection representation of the Regge action \cite{our3}. More accurately, realization of SO(3,1) as a subset in $SO(3,C) \times SO(3,C)$ was considered. The functional integration over connection in the factorization case $N = 0$ can be calculated in closed form; also typical expressions can be calculated entering further terms in the expansion in powers of $N$. These are exponentially suppressed at large areas and possess regular behavior at finite, in particular, at small areas.

The method used, starting with zero $N$, leads in zero order to the pure module of the result of the functional integration over connection; the phase of the result is contributed by further orders of expansion in powers of $N$. We expect that the phase is basically close to the Regge action. In favor of this is the fact that we can get some other expansion whose applicability conditions do not contradict to those of the considered expansion in powers of $N$ and such that in zero order we get just the phase, and this is exactly the Regge action. Namely, we can expand the action as a function of connection about the extremum point as a Taylor series in powers of a certain connection variable of the type of generator of connection (but not quite that and such that in terms of it the Haar measure would be some Lebesgue measure). The extremum is a solution of the classical equations of motion for connection, that is, the zero order term in this series is just the Regge action which thus gives the phase. The terms of order higher than second can, in principle, be considered as a perturbation with respect to the main part $S_0$ of the action $S$ consisting of the terms up to the second order. Integration of $\exp(iS)$ over connection reduces to integration of the monomials of the connection variables with $\exp(iS_0)$. This will be expansion in powers of $G$. Now, when we confine ourselves to the terms in the action up to a certain (the second) order in the connection variable, the belonging of the curvature matrices and parts of the action to the dependent or independent ones is preserved. In particular, $N$ enters as scales in the contribution of the dependent curvatures (on the t-like triangles), and, as above, we expect analyticity with respect to $N$ of the result of the integration over connection.

In the above mentioned case of the orthogonal connection representation, one else argument clearly illustrates regularity of the dependence on $N$, for simplicity, in the absence of the parity violation term. At $N = 0$ the bilinear form in $S_0$ becomes degenerate. However, in the considered case, the limits of integration over the connection variables are not all infinite, because connection is not infinitesimal, but finite, and can be an Euclidean rotation which is compact. $N$ is a coefficient in the contribution into action of the curvature on a timelike area. Qualitatively, we can say that the presence of $N \neq 0$ contributes to the convergence of integration over that part of this curvature which rotates around this area, that is, just over an Euclidean rotation. One can imagine that this rotation is temporarily used as some independent variable, the integration over which is compact and finite regardless of whether $N$ is zero or not. Thus, the result is not singular at $N = 0$.

Thus, there are some two expansions of the result of the functional integration over connection whose applicability conditions do not contradict to each other, leading in the zero order: one to some pure module, the other to some pure phase of the result. The phase is the Regge action $S_{\rm Regge}$, the module $F$ is evaluated for some limiting values of $\gamma$ at $N = 0$ for the case of the affine connection representation in the present work or calculated in our paper \cite{our3} for the case of the orthogonal connection representation for any $\gamma$ at $N = 0$ together with the typical expressions appearing in the expansion in powers of $N$. The result
\begin{equation}\label{FexpiS}                                             
F \exp (i S_{\rm Regge} ),
\end{equation}

\noindent already in terms of the pure edge (or area) variables, can be used to write out a generating functional for some diagram technique. The action $S_{\rm Regge}$ itself can lead to the perturbation theory diagrams, for example, for the graviton propagator and other Green functions. The graviton mode was considered in the work \cite{RocWil} for the simplest periodic Regge lattice. The system is similar to an ordinary lattice with spacings equal to the edge lengths, which are field variables over which functional integration with the weight $F$ is now performed.

Due to $F$, the diagram is averaged over elementary lengths/areas in the vicinity of Planck scale values. In the present case, this is most clearly demonstrated with the help of the obtained idealized (singular) delta-function form of $F$ which directly equate areas to Planck scale values (in a real situation, $F$ will be bell-shaped).

However, the effect can be seen also if $F$ is exponentially suppressed at large areas and is simply bounded as in the case of the orthogonal connection representation \cite{our3}, if we consider the measure in the tetrad variables to integrate in the resulting path integral with (\ref{FexpiS}). The full measure contains, in addition to $F$, also the metric or tetrad part of the original path integral measure for the connection representation. The path integral measure usually can be fixed in the Hamiltonian formalism ($\d p \d q$), in the considered case of the discrete system - after a temporary transition to the limit of a continuous time, when the edges are shrinking in the time direction. The conjugate variables (like $p$, $q$, but in more complex nonlinear way) in the continuous time limit of the orthogonal connection representation of Regge calculus are some (combinations with their dual) tensors $v^{ab}_{\sigma^2}$ of certain triangles $\sigma^2$ and some orthogonal connection matrices $\Omega_{\sigma^3}$ on certain tetrahedra, so that the measure symbolically is $\d v \D \Omega$, where $\D \Omega$ is the invariant (Haar) measure. Now we have a set of the local inertial frames, and the tensor of the same triangle $\sigma^2$ can be defined in the frame of each 4-simplex sharing it, $\sigma^4 \supset \sigma^2$, and denoted as $v^{ab}_{\sigma^2 | \sigma^4}$. Returning backward to the fully discrete theory, we need to consider, from symmetry considerations, taking the product of the volume elements $\d^6 v^{ab}_{\sigma^2 | \sigma^4}$ over all the leaf/diagonal $\sigma^2$s and $\sigma^4 \supset \sigma^2$. In this theory, it was generally natural to start from the general form of the tensors $v^{ab}$, which do not presuppose the existence of the edge vectors on which these tensors are constructed. Then the measure in such an extended configuration superspace is projected onto the hypersurface of true gravity by inserting $\delta$-function factors ensuring the existence of the edge vectors and metric for which these tensors are bivectors, and the unambiguity (continuity) of the metric induced on the 3-dimensional faces of the 4-simplices. These factors can be chosen invariant with respect to an arbitrary redefinition of the tetrahedron edges \cite{our4} (ie, existence and continuity of metric are purely local properties not depending on the form of the simplex with the help of which these are formulated). To summarize, at small areas the phase volume is suppressed due to a sufficient number of volume elements $\d^6 v^{ab}_{\sigma^2 | \sigma^4}$ together with the triangle inequalities for the edges which considerably limits the set of feasible configurations in the configuration superspace. At large areas, as mentioned, we have exponential suppression due to $F$. In the absence of a sufficient suppression at large areas, the pure Lebesgue measure on area tensors would lead to divergences in this region. That is, the suppression at small areas is possible due to the exponential suppression at large areas.

The situation looks similar to that in estimating the usual diagrams of perturbation theory by cutting off at Planck scales. (Meanwhile, a certain perturbation theory series, although in connection representation, has already been partially summed up to give a {\it nonperturbative} contribution to $F$.) But now this is not an estimate of the order of magnitude, but some precise calculation. Since the cut off parameter is the coupling constant $G$ itself, these series are, in fact, numerical ones. We expect that the loop smallness will play the role of a small parameter of the series and that the series are asymptotical as in QED. Further work consists in studying this finite diagram technique, including also finding and accounting for the corrections to the module $F$ and the phase $S_{\rm Regge}$ in the generating functional.

\section*{Acknowledgments}

The present work was supported in part by the Ministry of Education and Science of the Russian Federation.

\end{document}